\documentclass[aps,prl,twocolumn,floats,epsfig,showpacs]{revtex4}
\usepackage{bbm}
\usepackage{amssymb}
\usepackage{ifpdf,psfrag}
\usepackage{graphicx,epsfig}

\newcommand{\eq}{\begin{equation}}
\newcommand{\en}{\end{equation}}
\newcommand{\ear}{\begin{eqnarray}}
\newcommand{\rae}{\end{eqnarray}}

\newcommand{\la}{{\cal L}}

\newcommand{\tr}{{\rm tr}\,}

\newcommand{\p}{{\partial}}

\begin{document}
\title{Dirac-Born-Infeld action from spontaneous breakdown of 
Lorentz symmetry\\ in  brane-world scenarios}
\author{F.\ Gliozzi}
\affiliation{
 Dipartimento di Fisica Teorica, Universit\`a di Torino, and\\
 INFN, Sezione di 
Torino, P.\ Giuria 1, 10125 Torino, Italy }

\begin{abstract}
In whatever Lorentz invariant theory, the presence of extended d-dimensional 
objects inside a higher dimensional bulk space-time, like for instance D-branes in string theories, induces a spontaneous breakdown of the Poincar\'e invariance of the
 bulk; however the effective action describing these extended objects should still respect this larger invariance through  a non-linear 
realization of the full symmetry. Here the specific form of such a 
realization in the presence of an electromagnetic field is uncovered. The
Dirac-Born-Infeld type action for a D-brane turns out to be invariant under such transformations. Conversely it is explicitly demonstrated in some simple cases that  
the most general invariant action, as long as derivatives of the 
field strength (and second derivatives of the scalars)  can be neglected, is a linear combination of terms of 
the Dirac-Born-Infeld type with different scale parameters. The string 
result is obtained if one further assumes that the theory depends upon a 
single length scale.    
\end{abstract} 


\maketitle

Spontaneously broken symmetry is a central concept of modern theoretical 
physics. It plays a major role both in the physics of phase transitions and critical phenomena and in the theory of fundamental interactions. When the broken symmetry is continuous,  the resulting massless Nambu-Goldstone bosons 
\cite{gold,namb} are the natural degrees of freedom to be taken into account in the low-energy regime of the theory. For a compact internal symmetry group $G$ broken  to an arbitrary subgroup $H$ there are ${\rm dim}(G)-{\rm dim}(H)$ Goldstone bosons needed to restore in a non-linear way the invariance of the effective action under $G$ \cite{wein,ccwz}. Non-linear realizations of $G$  
tell us more about the Goldstone bosons than just they are massless; they also 
tightly constrain their interactions. When $G$ is a group of space-time 
transformations there is no longer a one-to-one correspondence between the number of broken generators and the number of Goldstone bosons, because only some of them are necessary to build a suitable non-linear realization of $G$. A famous example is the description of a relativistic bosonic string in a Minkowski 
space-time of dimension D in the light-cone gauge \cite{ggrt}, where a non-linear realization of the full D-dimensional  Poincar\'e group  is given in terms of the D-2 transverse coordinates $X_\perp$ of the string, which are the Goldstone modes associated with the spontaneously broken translational invariance in the transverse directions. A similar realization in the static gauge has been recently used 
\cite{aksc,oamf} to show, generalizing earlier ideas of Ref. \cite{luwe}, that the most general Lorentz-invariant string 
action written in terms of the $X_\perp$'s is 
the Nambu-Goto action \cite{ggrt}, as long as the second derivatives 
of the $X_\perp$'s can be neglected. 

The main purpose of this note is  to enlarge this kind of analysis  to a more general d-dimensional extended object characterized by the presence of a Maxwell field. 

More precisely we assume that there is a Poncar\'e-invariant theory in a D-dimensional Minkowski space-time admitting a stable solution with a 
d-dimensional extended object.  
Its position may be given by the coordinates $x_\mu$ with $\mu=0,1\dots, D-1$,
where $x_a\,(a=0,\dots,d-1)$ are the internal coordinates, while the remaining transverse coordinates  are functions of the position on the world-volume: $x_i=X_i(x_0,x_1,\dots,x_{d-1})$ with  $i=d,d+1,\dots,D-1$. They describe the embedding of the 
extended object in the space-time. Clearly this solution spontaneously breaks the transverse translational invariance along the $i$ directions 
as well as  the  Lorentz invariance $SO(1,D-1)$ of the bulk space-time, which is 
broken  to $SO(1,d-1)\times SO(D-d)$, whereas the action describing the full theory is kept untouched, of course. Thus, integrating out the heavy modes of the theory  leaves an effective d-dimensional action for the massless modes which should be invariant under the full $SO(1,D-1)$ Lorentz group. We assume that the massless modes   propagating through the extended object are the $D-d$ transverse 
scalar fields $X_i$ describing the fluctuations of the extended object (the Goldstone modes), and a d-dimensional gauge field $A_a$. Since these fields 
form incomplete  multiplets of $SO(1,D-1)$, they are forced to transform 
non-linearly under it.

How do these fields transform ? 
So far, an  answer has been found only for the $X_i$'s. 
Consider an infinitesimal $SO(1,D-1)$ transformation of parameter $\epsilon$ in the plane $(b\,i)$, with $0<b<d\le i< D$. We have \cite{ggrt,aksc,oamf}   
\eq
\delta^{bi}_\epsilon X_\mu=\epsilon\left(\delta_{b\mu}X_i-\delta_{i\mu}x_b-X_i
\frac{\partial X_\mu}{\partial x_b}\right)\,.
\label{bix}
\en
If $b$ is a temporal index, i.e. $b=0$, it suffices to change the sign of  the second term of the above expression. To simplify the notation 
we will consider only transformations involving  spatial indices. 
The recipe to write down Eq.(\ref{bix}) is very simple: the standard linear 
transformation which mixes coordinates and fields is followed by a reparametrization by which the new observer adjusts the description to his own world-volume 
frame. Here the non-linearity is trivially due to a particular choice of 
parametrization of the world-volume; other more symmetric choices are possible  where the whole Lorentz group is linearly realized \cite{ggrt}. 

What would seem  to be a much more difficult issue is the way of transforming of the gauge field, owing to the fact that this problem is presumably unrelated to the 
way of parameterizing the extended object. Actually it appears to be no 
 specific guiding principle, nevertheless a trial and error method 
 produced a surprisingly simple solution
\eq
\delta^{b i}_\epsilon A_a=-\epsilon\left(A_b\frac{\partial X_i}{\partial x_a}+
X_i\frac{\partial A_a}{\partial x_b}\right)\,.
\label{bia}
\en
The joint transformations (\ref{bix}) and (\ref{bia}) generate a non-linear realization of the whole $SO(1,D-1)$ group.
It reduces to a linear representation of the unbroken subgroup  $SO(1,d-1)\times SO(D-d)$ according to the symmetry breaking pattern.
In particular the
commutator $[ \delta^{ci}_\eta,\delta^{bi}_\epsilon]
\equiv \delta^{ci}_\eta\delta^{bi}_\epsilon-\delta^{bi}_\epsilon\delta^{ci}_\eta $, after several remarkable cancellations, becomes, as expected,
\eq
[ \delta^{ci}_\eta,\delta^{bi}_\epsilon]A_a=
\epsilon\eta\left(x_c\frac{\partial A_a}{\partial x_b}-
x_b\frac{\partial A_a}{\partial x_c}+\Sigma^{cb}_{a e}\, A^e\right),
\label{rota}
\en 
with  $\Sigma^{cb}_{a e}=\delta_{ac}\delta_{be}-\delta_{ab}\delta_{ce}$.
Similarly
\eq
(\delta^{b\,j}_\eta\delta^{b\,i}_\epsilon-\delta^{b\,i}_\epsilon\delta^{b\,j}_\eta)
A_a=0\,.
\en
In Eq. (\ref{rota}) and in the following it is implied that a summation 
has to be performed over any symbol appearing once as an upper and once as a 
lower index. 

Incidentally, we note that the proposed set of transformation rules for 
the gauge field is compatible with gauge invariance, in the sense that 
if $A_a$ is a pure gauge, i.e. $A_a=\frac{\p \Phi}{\p x_a}$, so also is
the transformed $A_a\to A_a'=A_a+\delta^{bi}_\epsilon A_a$, with a transformed function $\Phi'=\Phi+\delta^{bi}_\epsilon\Phi$, where
\eq
\delta_\epsilon^{bi}\Phi=-\epsilon X_i\frac{\p \Phi}{\p x_b}~.
\label{bis}
\en  
Clearly this equation also provides the transformation law of a scalar field propagating in the extended object. 

An important consequence of Eq.(\ref{bia}) is that
the way of transforming of the field strength $F_{ab}=\frac{\partial A_a}
{\partial x_b}- \frac{\partial A_b}{\partial x_a}$, namely 
\eq
\delta^{ci}_\epsilon F_{ab}=-\epsilon\left(
\frac{\partial X_i}{\partial x_a} F_{cb}+
\frac{\partial X_i}{\partial x_b} F_{ac}+X_i\frac\partial{\partial x_c}
F_{ab}\right)\,,
\label{bif}
\en
is exactly the same as that of the induced metric $g_{ab}$, i.e.
\eq
\delta^{ci}_\epsilon g_{ab}=-\epsilon\left(
\frac{\partial X_i}{\partial x_a} g_{cb}+
\frac{\partial X_i}{\partial x_b} g_{ac}+X_i\frac\partial{\partial x_c}
g_{ab}\right),
\label{big}
\en
with
\eq
g_{ab}=\frac{\partial X^\mu}{\partial x_a}\frac{\partial X_\mu}{\partial x_b}
\equiv\eta_{a\,b}+
\frac{\partial X^i}{\partial x_a}\frac{\partial X_i}{\partial x_b}=\eta_{ab}+
h_{ab}~,
\label{metric}
\en
where $\eta_{ab}$ is the diagonal Minkowski metric with 
$1=-\eta_{00}=\eta_{aa}~(a=1,\dots,d-1)$. It follows that one can
 take arbitrary linear combinations of $F_{ab}$ and $g_{ab}$ without 
altering their covariance properties under $SO(1,D-1)$. This will play a 
crucial role in the following. Our conventions are such that both
the gauge fields and the scalars  $X_i$ have the dimensions of length, hence 
both $g_{ab}$ and $F_{ab}$ are dimensionless quantities.

Gauge and $SO(D-d)$ invariance imply that the effective Lagrangian density 
$\la_{d,D}$ of the extended object is a function of $F_{ab}$ and $h_{ab}$ as well 
as of their derivatives, however we assume that the latter can be neglected, as 
customary in string calculations on this subject \cite{clny,leig}. Thus the 
problem we have to solve reduces to find the most general effective action
\eq  
S=T_d \int d^dx\,\la_{d,D}(F,h)\,,
 \en
such that $\delta^{bi}_\epsilon S=0$. $T_d$ is a parameter with the dimension of $length^{-d}$ which is called brane tension in string theory. It is further assumed that $\la_{d,D}$ 
can be Taylor-expanded in its arguments and that the integration 
domain is the whole d-dimensional Minkowski space, so we will not have to consider boundary terms.

Although a general proof is still lacking, we shall find considerable 
circumstantial 
evidence supporting the claim that the most general effective action fulfilling the above constraints is a linear combination of actions of Dirac-Born-Infeld type. 

More precisely, we will show that 
\emph{i)} the first few terms of the Taylor expansion  of $\la_{d,D}$ are uniquely determined by the requirement of Lorentz invariance generated by the non-linear transformations (\ref{bix}) and (\ref{bia}); \emph{ii)} these transformations 
generate  a set of recurrence relations for the Taylor coefficients that we solve explicitly in the case of a three-dimensional extended object embedded in a four-dimensional Minkowski space-time. We also solve the recurrence relations relative to the $X_i$ field for any d-dimensional extended object embedded in a 
(d+1)-dimensional space-time; the corresponding Taylor series can be written in a closed form and gives the desired result; \emph{iii)} finally we shall prove explicitly that the  general Dirac-Born-Infeld action in any dimension is invariant under 
(\ref{bix}) and (\ref{bia}).     

The first two points are a direct consequence of the way of transforming of the quantity $h_{ab}$ defined in Eq.(\ref{metric}). Actually, at variance with the 
transformation law of $g_{ab}$, which is homogeneous, $h_{ab}$ contributes with an inhomogeneous term which lowers the degree of the powers of $h$. If we use the short-hand notation $\delta^{ci}_\epsilon g_{ab}=A[^{ci}_\epsilon]^{ef}_{ab}\,g_{ef}$ for the homogeneous transformation (\ref{big}), we have
\eq
\delta^{ci}_\epsilon h_{ab}=A[^{ci}_\epsilon]^{ef}_{ab}\,h_{ef}-\epsilon
\left(\frac{\partial X_i}{\partial x_b}\delta_{ac}+
\frac{\partial X_i}{\partial x_a}\delta_{cb}\right)~.
\label{bih}
\en    
 In the  Taylor expansion of $\la_{d,D}$, the 0-order terms in $F$  
 are 
\eq
\la^{(0)}_{d,D}=-\frac12 \tr h+\alpha_1 (\tr h)^2+
\alpha_2 \tr h^2+O(h^3)\,,
\label{la0}
\en
where 
\eq
\tr h^n=\eta^{a_1b_1}\eta^{a_2b_2}\dots \eta^{a_nb_n}h_{b_na_1}h_{b_1a_2}\dots  
h_{b_{n-1}a_n}\,.
\en 
Eq.(\ref{bih}) yields, up to $O(h^2)$ terms and a total derivative, 
\eq
\delta^{bi}_\epsilon \la^{(0)}_{d,D}/\epsilon= -\frac{8\alpha_1+1}2\frac{\p X_i}{\p x_b}
\tr h-(4\alpha_2-1)h_b^a\frac{\p X_i}{\p x_a}\,.
\en
 Demanding this result to vanish fixes the values of $\alpha_1$ 
and $\alpha_2$, as first observed in \cite{oamf}. Similarly,  the first 
contributions of the Maxwell field are
\eq
\la^{(2)}_{d,D}=\frac14 \tr F^2+\beta_1\tr F^2 \tr h+\beta_2 \tr(hF^2)+O(h^2).
\en
They give, up to $O(h)$ terms,
\eq
\delta^{bi}_\epsilon \frac{\la^{(2)}_{d,D}}\epsilon=(2\beta_2+1)\frac{\p X_i}{\p x_c}
F_{be}F^{ce}-\frac{8\beta_1-1}4 \frac{\p X_i}{\p  x_b} \tr F^2.
\en
Again, demanding this result to vanish now fixes the coefficients
 $\beta_1$ and $\beta_2$.

It is almost evident that one can enlarge this analysis to the whole Taylor expansion in $h$. The most general invariant term contributing to the 
$n^{th}$ order 
terms of $\la_{d,D}^{(0)}$ can be written in the form of a multi-trace
\eq
(\tr h)^{n_1}(\tr h^2)^{n_2}\dots(\tr h^d)^{n_d}\,,~\sum_{k=1}^dk\,n_k=n\,.
\en  
In the general case the explicit form of the recurrence relations is rather involved, but when the dimension D of  space-time is 
${\rm D=d+1}$ it simplifies dramatically. In this case we have $\tr h^k=(\tr h)^k$ for any integer $k$, so
 Eq. (\ref{la0}) and the consequent recurrence relations become simply
\eq
\la_{d,d+1}^{(0)}=\sum_{k=0}^\infty c_k (\tr h)^k\,;~k\,c_k+(k-\frac32)\,
c_{k-1}=0\,.
\en
The solution with initial condition $c_1=-\frac12$ is  the binomial
$c_k=-\left(\matrix{\frac12\cr k}\right)$, therefore
we obtain, as expected,
\eq
\la_{d,d+1}^{(0)}=-\sqrt{1+\tr h}=-\sqrt{-\det{(\eta_{ab}+h_{ab})}}~.
\en
Switching on the gauge 
field introduces in the Lagrangian density $\la_{d,D}$ several 
independent single-trace mixed invariants of the form
\eq
\tr (F^{m_1} h^{n_1}F^{m_2}h^{n_2}\dots)\,;
\en
when ${\rm D=d+1}$ they split into multi-trace terms of the kind
\eq
\tr(F^{m_1}h)\tr(F^{m_2}h)
\tr(h^{n_1-1})\tr(h^{n_2-1})\dots\,.
\en
To make the calculation explicit and simple, we specialize now to the case
${\rm d=3}$ and ${\rm D=4}$. In this case we can expand the most general 
Lagrangian density in terms of only three invariants
\eq
\la_{3,4}=\sum_{n=0}^\infty\sum_{p=0}^\infty\sum_{q=0}^\infty c_{n,p,q}(\tr h)^n
(\tr F^2)^p(\tr hF^2)^q\,.
\label{lat}
\en  
Lorentz invariance of the action with respect to the joint 
transformations (\ref{bif}) and (\ref{bih}) dictates the form of the 
recurrence relations among the $c_{n,p,q}$'s. We find
\eq
n\,c_{n,p,q}+(q+n-\frac32)\,
c_{n-1,p,q}=0\,,
\label{reca}
\en
and
\eq
2p\,c_{n,p,q-1}+
q\, c_{n-1,p-1,q}+ q\, c_{n,p-1,q}    =0\,.
\label{recb}
\en
The general solution of (\ref{reca}) is
\eq
c_{n,p,q}=\left(\matrix{\frac12-q\cr n\cr}\right)\,\psi(p,q)\,,
\en  
where $\psi$ is arbitrary. We can go one further step by inserting this expression in the second set of recurrence relations. It gives
\eq
c_{n,p,q}=\frac{-8^{-p}}{1-2p}\left(\matrix{\frac12-q\cr n\cr}\right)
\left(\matrix{\frac12-p\cr q\cr}\right)\left(\matrix{2p\cr p\cr}\right)
\phi(p+q).
\en 
The arbitrary function $\phi$ fixes the relative scale of the different powers of $F$ in the action.  If we choose $\phi(p+q)=\lambda^{p+q}$ the Taylor expansion (\ref{lat}) can be resummed,  yielding
\eq
\la_{3,4}=-\sqrt{\left(1+\tr h\right)\left(1-\lambda\frac{\tr F^2}2\right)
+\lambda\frac{\tr(h F^2)}2}\,,
\en
which can be easily rewritten in the Dirac-Born-Infeld form
\eq
\la_{d,D,\lambda}=
-\sqrt{-\det\left(g_{ab}+\lambda\,F_{ab}\right)}\,.
\label{action}
\en
The most general solution of the above recurrence relations  
reads
\eq
\la_{3,4}=\sum_kc_k\la_{3,4,\lambda_k}~;~\sum_kc_k=1\,.
\en
where the arbitrary  parameters $\lambda_k$ reconstruct the function 
$\phi(p+q)$. 

We are now in a position to provide a simple proof of the invariance of the 
Dirac-Born-Infeld action $S=T_d\int d^dx\la_{d,D,\lambda}$ in its full generality,  with no restrictions on the space-time dimensions ${\rm d<D}$. Let $f(e)$ be an arbitrary function of the two-index tensor
\eq
e_{ab}=g_{ab}+\lambda F_{ab}~.
\en
The joint transformations (\ref{bif}) and (\ref{big}) give
\eq
\delta^{ci}_\epsilon f(e)=-\epsilon \Lambda_c^a\frac{\p X_i}{\p x_a}-\epsilon
X_i\frac{d}{d x_c}f(e)\,,
\label{cif}
\en
with
\eq
\Lambda_c^a= \frac{\p f}{\p e_{a'b}}e_{cb'}\eta^{bb'}\eta^{a'a}+
\frac{\p f}{\p e_{ba'}}e_{b'c}\eta^{bb'}\eta^{a'a}~.
\en 
If $f(e)$ is chosen in such a way that the d-dimensional 
gradient 
$\frac{\p X_i }{\p x_a}$ is an eigenvector of $\Lambda$ corresponding to 
the eigenvalue $f(e)$, then (\ref{cif}) becomes a total derivative. This is precisely what happens if we set $f(e)=\sqrt{-\det(e_{ab})}$. In fact, putting
$\delta e_{ab}=e_{cb}\frac{\p X_i}{\p x_a}+e_{ac} \frac{\p X_i}{\p x_b} $, 
standard manipulations of the determinant yield
\eq
\delta\det(e)=\det(e)e^{ba}\delta e_{ab}=2\det(e)\frac{\p X_i}{\p x_c}~,\, 
\label{det}
\en
where $e^{ab}$ is the matrix inverse of $e_{ab}$ with 
$e^{ab'}e_{b'c}=e^{b'a}e_{cb'}=\delta^a_c$. Eq. (\ref{det}) tells us that  
the $\Lambda$ associated with $\det(e)$ has eigenvalue 
$2\det(e)$, so we have at once
\eq
\delta^{ci}_\epsilon\la_{d,D,\lambda}=
\epsilon \frac\p{\p x_c}\left( X_i \sqrt{-\det(g_{ab}+\lambda F_{ab})}\right)\,,
\label{invar}
\en
which is our concluding result. It states that the Dirac-Born-Infeld type action
$T_d\int d^dx\la_{d,D,\lambda}$ in arbitrary space-time dimensions is invariant with respect to the infinitesimal non-linear transformations (\ref{bif}) and 
(\ref{big}) generating the Lorentz group of a higher dimensional space.

One may ask why the simple and general invariance   of 
the Dirac-Born-Infeld  action we described in this note has apparently 
not been noticed before. A 
possible answer is that it has not been realized, so far,  that in the 
spontaneous breakdown of the Lorentz symmetry of the bulk space associated with  brane formation not only the Goldstone modes but all the massless degrees of freedom propagating in the brane should transform non-linearly under the 
full Lorentz group. In the string approach to D-brane action \cite{clny,leig} there is another massless bosonic field  which has to be considered, an antisymmetric tensor $B_{ab}$ that, by consistency, should transform non-linearly like $g_{ab}$ and $F_{ab}$. It would be interesting to try to extend our considerations to the case of $N$ coincident D-branes, where the world-volume theory involves a $U(N)$ gauge theory \cite{aats,myer}. It would be also very interesting if one could say something about couplings that involve derivatives of the field
strength and the second derivative of the scalars, generalizing the analysis 
of Ref. \cite{oamf}. 

It is important to point out that the string derivations of the Dirac-Born-Infeld action and its generalizations make use of quantum properties of the underlying string theory,  e. g. the vanishing of the open string $\beta$-functions 
\cite{leig} or T-duality arguments \cite{myer}, whereas in the present 
approach we considered only symmetry properties of the classical action. At the quantum level the non-linear generators 
of the full  Lorentz invariance are presumably anomalous. 
For instance in the Nambu-Goto string in the light-cone gauge as well as in 
the static gauge  the transformation (\ref{bix}) 
of the $X_i$'s  is anomalous unless D=26 \cite{ggrt}. 
In the superstring approach,  one would presumably find similar Lorentz 
anomalies unless all the massless bosonic and fermionic degrees of freedom 
of the spectrum \cite{gso}  are taken into account and D is 
set to its critical value. In this case there is strong evidence that the effective four-dimensional action of noncommutative gauge theory - used to describe D-branes with B-field background - does indeed enjoy the full $SO(1,9)$ symmetry \cite{Blaschke:2010rr}.

The idea of this work originated during the workshop ``Confining flux tubes and strings''  held in ECT*, Trento 5-9 July 2010. The author would like to thank the organizers and all the participants for fruitful, stimulating discussions. 


\begin{thebibliography}{20}
\bibitem{gold} J. Goldstone, Il Nuovo Cimento {\bf 19}, 154 (1961).
\bibitem{namb} Y. Nambu, Phys. Rev. Lett. {\bf 4}, 380 (1960). 
\bibitem{wein} S. Weinberg, Phys. Rev. Lett. {\bf 18}, 188 (1967).
\bibitem{ccwz} S. Coleman, J. Wess and B. Zumino, 
Phys. Rev. {\bf 177}, 2239 (1969); C. G. Callan, S. Coleman, 
J. Wess and B. Zumino, Phys. Rev. {\bf 177}, 2247 (1969).
\bibitem{ggrt} P. Goddard, J. Goldstone, C. Rebbi and C.B. Thorn,
Nucl. Phys. B {\bf 56 }, 109 (1973).
\bibitem{aksc} O. Aharony, Z. Komargodski and  A. Schwimmer, work in progress, presented by O. Aharony at the String 2009 conference, June 2009 and at 
ECT* workshop on `` Confining flux tubes and strings'', July 2010.
\bibitem{oamf} O. Aharony and M. Field, 
  JHEP {\bf 1101}, 065 (2011) [arXiv:1008.2636 [hep-th]].
\bibitem{luwe} M. L\"uscher and P. Weisz, JHEP {\bf 0407}, 014 (2004) 
[arXiv:hep-th/040605].
\bibitem{clny} C.G. Callan, C.Lovelace, C. R. Nappi and S. A. Yost, 
Nucl. Phys. B {\bf 308},221 (1988); A. Abouelsaood, C.G. Callan, C. R. Nappi and S. A. Yost, Nucl. Phys. B {\bf 280}, 599 (1987).
\bibitem{leig} R.G. Leigh, Mod. Phys. Lett. {\bf A4}, 2767 (1989).
\bibitem{aats} A.A. Tseytlin, Nucl. Phys. B {\bf 501}, 41 (1997)
[arXiv:hep-th/9701125].
\bibitem{myer} R.~C.~Myers, 
  JHEP {\bf 9912}, 022 (1999) [arXiv:hep-th/9910053].
\bibitem{gso}
  F.~Gliozzi, J.~Scherk and D.~I.~Olive,
  Phys.\ Lett.\  B {\bf 65}, 282 (1976);
  Nucl.\ Phys.\  B {\bf 122}, 253 (1977).
\bibitem{Blaschke:2010rr}
  D.~N.~Blaschke, H.~Steinacker and M.~Wohlgenannt,
  JHEP {\bf 1103}, 002 (2011)
  [arXiv:1012.4344 [hep-th]].
\end{thebibliography}
\end{document}